\documentclass[twocolumn]{aastex63}
\usepackage{mathrsfs}
\usepackage[]{txfonts}
\usepackage{comment}
\submitjournal{AJ}
\accepted{April, 4, 2021}

\shortauthors{Miyakawa et al.}
\shorttitle{Multicolor Photometry}

\begin{document}
\title{Joint Analysis of Multicolor Photometry: 
A New Approach to Constrain the Nature of Multiple-Star Systems 
Hosting Exoplanet Candidates}

\def\myemail{miyakawa.k.aa@m.titech.ac.jp}
\def\titech{Department of Earth and Planetary Sciences, Tokyo Institute of Technology, Meguro-ku, Tokyo, 152-8551, Japan}
\def\tsinghua{Tsinghua Center for Astrophysics, Tsinghua University, Beijing 100084, China}
\def\oao{Okayama Astrophysical Observatory, National Astronomical Observatory of Japan, Asakuchi, Okayama 719-0232, Japan}
\def\naoj{National Astronomical Observatory of Japan, NINS, 2-21-1 Osawa, Mitaka, Tokyo 181-8588, Japan}
\def\tokyo{Department of Astronomy, Graduate School of Science, The University of Tokyo, Hongo 7-3-1, Bunkyo-ku, Tokyo, 113-0033, Japan}
\def\abc{Astrobiology Center, NINS, 2-21-1 Osawa, Mitaka, Tokyo 181-8588, Japan}

\def\komaba{Komaba Institute for Science, The University of Tokyo, 3-8-1 Komaba, Meguro, Tokyo 153-8902, Japan}
\def\JST{JST, PRESTO, 3-8-1 Komaba, Meguro, Tokyo 153-8902, Japan}
\def\abc{Astrobiology Center, 2-21-1 Osawa, Mitaka, Tokyo 181-8588, Japan}
\def\IAC{Instituto de Astrof\'{i}sica de Canarias (IAC), 38205 La Laguna, Tenerife, Spain}

\author{Kohei Miyakawa}\affiliation{\titech}
\author[0000-0003-3618-7535]{Teruyuki Hirano}\affiliation{\titech}
\author{Bun'ei Sato}\affiliation{\titech}
  
\author{Akihiko Fukui}\affiliation{\tokyo}\affiliation{\IAC}
\author[0000-0001-8511-2981]{Norio Narita}\affiliation{\IAC}\affiliation{\abc}\affiliation{\komaba}\affiliation{\JST}
\correspondingauthor{Kohei Miyakawa}
\email{miyakawa.k.aa@m.titech.ac.jp}
\begin{abstract}
 We present a new method to assess 
 the properties of transiting planet candidates by multicolor photometry.
 By analyzing multicolor transit/eclipse light curves and apparent magnitudes
 of the target in parallel, this method attempts to identify the nature of the system and 
 provide a quantitative constraint on the properties of unresolved companion(s). 
  We demonstrate our method by observing the six systems hosting candidate transiting planets, identified by the K2 mission (EPIC 206036749, EPIC 206500801, EPIC 210513446, EPIC 211800191, EPIC 220621087, and EPIC 220696233). 
  Applying our analysis code to the six targets, we find that
  EPIC 206036749, EPIC 210513446, and EPIC 211800191 are likely to be triple-star systems including eclipsing binaries,
  and EPIC 220696233 is likely a planetary system, 
  albeit further observations are required to confirm the nature.
  Additionally, we confirm that the systematic errors in the derived system parameters arising from adopting specific isochrone models and observing instruments (passbands) are relatively small.
  While this approach alone is not powerful enough to validate or refute planet candidates,
  the technique allows us to constrain 
  the properties of resolved/unresolved companions, and prioritize the planet candidates for further follow-up observations (e.g., radial-velocity measurements). 
 
\end{abstract}


\section{Introduction}\label{sec:intro}
Dedicated space missions like Kepler, K2 \citep{Borucki2010, Howell2014},  CoRoT \citep{Auvergne2009}, and TESS \citep{Ricker2015} have unveiled 
detailed properties of transiting exoplanets.
While those survey missions can detect a large number of planetary candidates,
it is not straightforward to distinguish so-called false positives (FPs) from 
true planets.
Targets including multiple-stars such as eclipsing binaries are often unresolved 
in the raw images due to large pixel sizes of the detectors \citep{Ciardi2015}.
Radial velocity (RV) measurements are the most reliable follow-up approach to identify 
true planetary systems.
However, it is time-consuming to follow up a large number of planetary candidates identified by the space missions.

In the last decade, a vetting technique using multiple observables
and statistical assessment (so-called ``validation")
has been developed and applied to constrain the nature of the systems hosting 
planet candidates \citep[e.g.,][]{Morton2015, Crossfield2016, Hirano2016, Livingston2018}.
The main concept of the validation is to ``statistically" rule out the scenario 
that the observed transit-like signals are caused by non-planetary phenomena including
stellar eclipses, based on the stellar population models in the Galaxy. 
For instance, \texttt{vespa} \citep{Morton2012, Morton2015} is one of the most powerful tools for the statistical planet validation, enabling us to efficiently discuss the 
statistical properties of transiting planets. 
One caveat in this methodology is, however, that such tools are not suitable for exploring uncommon targets because it relies on the statistics and population model of binary or triple-star systems \citep{Girardi2005},
suggesting that true planetary systems are sometimes    mis-classified 
as false positive with a certain probability. 
In particular, this mis-classification is expected to happen when the target has an unresolved stellar companion. 
Planets in binary (or triple-star) systems have been detected and discussed in the literature \citep[e.g.,][]{Ngo2017,Matson2018},
but most of those planets were discovered around the brighter (brightest) star in 
each system, and there have been a very small number of planets orbiting the
``fainter" star in a multiple-star systems \footnote{https://www.univie.ac.at/adg/schwarz/multiple.html}. 
While transiting planets around the fainter ones in multiple-star systems generally
produce weaker observational signals, some of those planets may well be overlooked
in the past, or mis-classified as false positives due to stellar multiplicity.

One method to constrain the nature of planet-candidate hosting targets without
resorting to the statistical models is to
combine ``multicolor" transit light curves with apparent magnitudes of the targets identified as candidate-planet hosts. 
These observables are both affected by the presence of the contaminant stars inside the photometric aperture.
Multicolor transit photometry has already been used in the validation of the planetary candidates \citep[e.g.,][]{Colon2011,Colon2012,Tingley2014, Parviainen2020, Parviainen2021}.
For example, the CoRoT space telescope performed the vetting with the three-color simultaneous photometry. 
This technique can identify the presence of contamination by detecting variations in transit depth and shape for different observing passbands.
Multicolor transit photometry is observationally ``cheaper" than 
RV measurements, since those observations are carried out with only $1-2$-m class telescopes.
\citet{Parviainen2019} presents an analytical approach for the multicolor transit light curves, which allows us to estimate the amount of flux contamination and true radius ratio for transiting candidates. 
More recently, \citet{Louie2020} conducted simulations of the detectability of transit depth variations among different passbands for TESS candidates.
%
A second set of photometric observables, the apparent magnitudes in different passbands, also have color information including the contaminations from unresolved stellar companion(s).
In \citet{Windemuth2019}, a method to estimate the stellar properties of eclipsing 
binaries without RV measurements by combining Kepler light curves 
with the spectral energy distribution (SED) based on apparent magnitudes was introduced.
They applied the method to the 728 Kepler eclipsing binaries,
and demonstrated the effectiveness of the SED correction.

In this paper, we present a new method, which combines both of the photometric observables (multicolor light curves and apparent magnitudes) and derives a possible nature of the candidate systems based on stellar isochrone models.
Our aim is to constrain physical properties (e.g., stellar mass and impact parameter, etc.) of the candidates without any assumptions on the underlying stellar populations as used in planet-validation tools (e.g., \texttt{vespa}). 
The concept to use the multicolor photometry is somewhat similar to the approach by  \citet{Parviainen2019}, but they did not attempt to directly constrain the physical properties of the contaminating sources.
We have developed a tool for these analyses, which helps us conduct subsequent follow-up observations efficiently.

The rest of this paper is organized as follows.
In Section \ref{sec:method}, we describe our photometric models 
and analysis procedure for planetary candidates.
We introduce our targets for demonstrations of our method and report the follow-up observations including multicolor transit photometry in Section \ref{sec:observation}.
Section \ref{sec:results} presents the results of the applications.
We discuss interpretations of the results and systematic effects in Section \ref{sec:discussion}.
Finally, Section \ref{sec:conclusion} is devoted to the summary and conclusion.

\begin{table*}[t]
\centering
\caption{\small Stellar parameters}\label{tab:stellar_param}
    \small
  \begin{tabular}{ccccrrrrc} \hline \hline
    EPIC ID & AO contaminant & $M_{\star}$($M_{\odot}$) & $R_{\star}$($R_{\odot}$) 
    	& $T_{\rm eff}$(K) & [Fe/H](dex) & $\log g $(dex) & parallax(mas)  & method \\ \hline 
    $206036749$ &negative & $1.07\pm0.05$ & $1.15\pm0.19$ & $5423\pm110$ & $+0.35\pm0.08$ 
    & $4.34\pm0.13$ & $~2.28\pm 0.05$& \texttt{SpecMatch-Emp} \\
    
    $206500801$ &no data & $1.17\pm0.06$ & $1.35\pm0.22$ & $5607\pm110$ &  $+0.38\pm0.08$ 
    & $4.47\pm0.12$ & $~2.48\pm 0.10$ &  \texttt{SpecMatch-Emp} \\
    
    $210513446$ &$\Delta {\rm mag_{K} \approx 1}$ at $0''.5$  & $0.91\pm0.04$ & $0.93\pm0.15$ & $5671\pm110$ & $-0.49\pm0.08$ 
    & $4.47\pm0.13$ & $~3.12\pm0.15$ & \texttt{SpecMatch-Emp} \\
    
    $211800191$ 	&negative & $1.05\pm0.05$ & $1.05\pm0.21$ & $5970\pm110$ & $-0.46\pm0.08$ 
    & $4.26\pm0.12$ & $~2.47 \pm 0.04$ &  \texttt{SpecMatch-Emp} \\
    
    $220621087$ & negative& $0.44\pm0.05$ & $0.43\pm0.04$ & $3585\pm{~~70}$ & $-0.32\pm0.12$ 
    & $4.86\pm0.04$ & $14.36 \pm 0.03$& \texttt{SpecMatch-Emp} \\
    
    $220696233$& negative& $0.58\pm0.07$ & $0.56\pm0.08$ & $3841\pm{~~76}$ & $+0.12\pm0.15$ 
    & $4.86\pm0.03$ & $~3.72 \pm 0.05$  &\citet{Mann2015}\\ \hline
  \end{tabular}

\end{table*}

\section{Method}\label{sec:method}
The observational imprints by the presence of contaminants on the multicolor transit light curves and SED should be treated in parallel, because they originate from the same sources.
In this section, we show how combining those pieces of information helps us derive the ``true nature"
(physical properties) of the candidate transiting systems.
We will first introduce the multicolor transit light curve model and SED respectively,
and describe the possible scenarios for the candidates and fitting process of the models.

\subsection{Multicolor Transit Photometry}
Multicolor transit photometry is a technique to observe planetary transits in multiple wavelength passbands \citep{Tingley2004, Colon2012}.
If another source contaminates in a photometric aperture, the observed light curve is diluted.
Due to the stellar color, the dilution ratio depends on each observing passband.
Hence, we can obtain the information about contaminants from multicolor transit photometry.
This technique is known as one of the planet-validation methods \citep[e.g.,][]{Parviainen2019}. 
The model of normalized light curve in a specific passband ($\lambda$) for a multiple-star system ($F_{\lambda}$) is expressed as
\begin{eqnarray}
F_{\lambda} = (f_{\lambda} - 1) \times d_{\lambda} + 1,
\end{eqnarray}
where $f_{\lambda}$ is the undiluted transit light curve in a specific passband whose analytic expression can be found, e.g., in \citet{Ohta2009},
and $d_{\lambda}$ is a dilution factor that is the flux ratio of the central star to the total system
including the companion(s).
Throughout this paper, we assume physically associated and unresolved targets (e.g., eclipsing binary, hierarchical triple binary or planet in a binary system).
In other words, we do not assume the contaminating light from background stars.
This is because, except for very rare cases of ``chance alignment," 
those background/foreground sources generally have a good spatial separation 
(e.g.,  $ > 0\farcs5$) from the target star, which can be easily identified by 
e.g., adaptive-optics (AO) or speckle observations.
Therefore, we can derive $d_{\lambda}$ using the following equation.
\begin{eqnarray}
\label{eq:2}
d_{\lambda} = \frac{1}{\sum_{i}10^{-\frac{2}{5}({\rm Mag}_{i, \lambda} - {\rm Mag}_{{\rm cen}, \lambda})}} , 
\end{eqnarray}
where ${\rm Mag}_{*, \lambda}$ denotes the absolute magnitude in a given passband $\lambda$, and
$i$ is the index for each star and ``cen" means central star,
which is the central star of the transiting/eclipsing system. 
In the calculation of the denominator, ``cen" is included in the index $i$.
To derive $d_{\lambda}$, we use the MESA Isochrones and Stellar Tracks\footnote{http://mesa.sourceforge.net/} \citep[MIST:][]{Dotter2016, Choi2016}
for the fiducial case, which covers a wide range of stellar effective temperature and metallicity for the synthetic absolute magnitude $\rm Mag_{\lambda}$. 
Later, we will also use other isochrone models to discuss the isochrone dependence of the fitting results. 

\subsection{SED}\label{ss:sed}
The shape of an observed SED also depends on stellar multiplicity.
We aim to accurately estimate the properties of the systems showing transit-like signals 
by adding SED information to multicolor transit light curves.
Here, we show our model of apparent magnitudes of the targets.
The apparent magnitude in a specific passband is described as
\begin{eqnarray}
{\rm mag}_{\lambda} = {\rm Mag}_{\lambda} + \mu + A_{\lambda},
\end{eqnarray}
where $\mu$ and $A_{\lambda}$ are the distance and the line-of-sight extinction modulus, respectively.
Since we assume unresolved multiple-star systems, 
${\rm Mag}_\lambda$ is obtained as total brightness of the stars in the system:
\begin{eqnarray}
{\rm Mag}_{\lambda} = -2.5 \log_{10}{\frac{1}{d_\lambda}} + {\rm Mag_{{\rm cen}, \lambda}},
\end{eqnarray}
where $d_{\lambda}$ is same variable in Equation (\ref{eq:2}).
We use the parallax from $Gaia$ DR2\footnote{https://gea.esac.esa.int/archive/} \citep{Gaia2016, Rene2018, Gaia2018b} for the distance modulus.
We multiply the color excess ($E(B-V)$) by empirical extinction vectors presented in \citet{Yuan2013} to derive $A_{\lambda}$,
where $E(B-V)$ is calculated with ${\rm Mag}_{B}$ and ${\rm Mag}_{V}$,
and the observational quantities ${\rm mag}_{B}$ and ${\rm mag}_{V}$. 
Here, we set the models of ${\rm mag}_{B}$ and ${\rm mag}_{V}$ to the observed values to derive $E(B-V)$.
We also apply the MIST isochrones for synthetic absolute magnitudes.
The apparent magnitudes in the $B$, $V$ \citep{Bessell1990}, $g$, $r$, $i$ \citep[Sloan Digital Sky Survey ;][]{York2000}, $J$,$H$,$K$ \citep[Two Micro All Sky Survey ;][]{Skrutskie2006}, W1, W2, W3 and W4 \citep[Wide-field Infrared Survey Explorer ;][]{Wright2010} bands are collected from catalogues on the VizieR\footnote{https://vizier.u-strasbg.fr/viz-bin/VizieR/} website. 
Finally, we calculate the synthetic apparent magnitudes in the passbands except $W3$ and $W4$, 
since \citet{Yuan2013} do not provide the extinction coefficients for those bands due to their low sensitivities.
\begin{table*}[t]
\centering
\caption{\small Stellar magnitudes}\label{tab:stellar_mgnitudes}
  \begin{tabular}{cccccccccc} \hline \hline
    band pass & 206036749 & 206500801 & 210513446 & 211800191 & 220621087 & 220696233&\\ \hline
    $B$ 	& $13.98 \pm 0.03$ 	& $12.44 \pm 0.27$	& $14.95 \pm 0.03$ 	
    		& $13.13 \pm 0.04$	& $15.66 \pm 0.15$	& $17.72 \pm 0.20$\\
    $g$ 	& $13.55 \pm 0.01$	& $12.55 \pm 0.01$	& $14.46 \pm 0.04$
    		& $12.80 \pm 0.06$ 	& $15.07 \pm 0.12$	& $17.12 \pm 0.04$\\
    $V$ 	& $13.21 \pm 0.02$	& $12.18 \pm 0.28$	& $13.98 \pm 0.07$
    		& $12.62 \pm 0.06$ 	& $14.14 \pm 0.04$	& $16.22 \pm 0.04$\\
    $r$ 	& $12.98 \pm 0.01$	& $11.96 \pm 0.01$	& $13.63 \pm 0.02$
    		& $12.42 \pm 0.02$ 	& $13.57 \pm 0.01$	& $15.76 \pm 0.04$\\
    $i$ 	& $12.78 \pm 0.02$	& $12.03 \pm 0.38$	& $13.26 \pm 0.04$
    		& $12.29 \pm 0.05$ 	& $12.66 \pm 0.02$	& $14.86 \pm 0.07$\\
    $J$ 	& $11.74 \pm 0.03$	& $10.75 \pm 0.02$	& $11.88 \pm 0.02$
    		& $11.37 \pm 0.02$	& $10.93 \pm 0.03$	& $13.15 \pm 0.03$\\
    $H$ 	& $11.38 \pm 0.03$	& $10.40 \pm 0.02$	& $11.34 \pm 0.03$
    		& $11.04 \pm 0.02$ 	& $10.34 \pm 0.02$ 	& $12.44 \pm 0.02$\\
    $K$ 	& $11.32 \pm 0.03$	& $10.30 \pm 0.02$	& $11.19 \pm 0.02$
    		& $10.96 \pm 0.02$	& $10.12 \pm 0.02$	& $12.29 \pm 0.02$\\
    $W1$ & $11.23 \pm 0.02$	& $10.24 \pm 0.02$	& $11.12 \pm 0.02$
    		& $10.94 \pm 0.02$	& $9.97 \pm 0.02$	& $12.18 \pm 0.02$\\
    $W2$ & $11.27 \pm 0.02$	& $10.28\pm 0.02$	& $11.10 \pm 0.02$
    		& $10.96 \pm 0.02$	& $9.88 \pm 0.02$	& $12.18 \pm 0.02$\\
     \hline
  \end{tabular}
\end{table*}
\begin{table}
\centering
\caption{\small Orbital parameters in original format}\label{tab:orbit_param}
  \begin{tabular}{cccc} \hline \hline
    EPIC ID& Period [day] & Depth [\%] & $R_{p}/R_{s}$\\ \hline
    $206036749^{(1)}$ & $1.131316_{\pm0.000030}$ & - &$0.0313_{\pm 0.0018}$\\
    $206500801^{(2)}$ & $8.15307$ & $2.292_{\pm 0.038}$ & - \\
    $210513446^{(1)}$ & $1.1489833_{\pm0.0000089}$ & - & $0.36_{\pm 0.17}$\\
    $211800191^{(2)}$ & $1.10605213$ & $0.147_{\pm 0.020}$ & -\\
    $220621087^{(3)}$ & $ 3.835582_{\pm0.000023}$& - & $0.0289_{-0.0010}^{+0.0019}$ \\
    $220696233^{(4)}$ & $28.735960_{-0.001521}^{+0.001529} $& - & $0.1056_{-0.0036}^{+0.0055}$\\ \hline
  \end{tabular}
  \begin{flushleft}
  	{\bf References :} (1) :  \citet{Crossfield2016}, (2): \citet{Barros2016}, 
	(3): \citet{Hirano2018a}, (4) : \citet{Livingston2018}.
   \end{flushleft}

\end{table}
\subsection{Fitting Process}
Our aim is to constrain the nature of the unresolved candidates ``quantitatively" 
by combining multicolor transit/eclipse photometry, SED, and isochrone fitting.
Specifically, we assume the following four scenarios for the systems hosting 
transiting-planet candidates:
\begin{enumerate}
\item planet (or dark object) transiting a single star, [PS];
\item planet (or dark object) transiting a star in a binary system, [PB];
\item eclipsing binary, [EB];
\item triple-star system including an eclipsing binary, [ET].
\end{enumerate}
The first two models are computed by setting the flux of the eclipsing object to 0.
Additionally, all models have no assumption on the hierarchical structure 
for the components (objects) within the system;
This means that the second model includes both p-type and s-type orbits 
and the latter two models include ``secondary eclipse" for the identified (primary) flux decrease.
In this paper, we do not consider systems with four or more stars, 
because such systems are much less common \citep[e.g., ][]{Raghavan2010} than 
the above cases.

We model the observed light curves ($F_{\rm\lambda}$) and apparent magnitudes ($\rm mag_{\rm\lambda}$) for each scenario using the values of $\rm Mag_{\rm \lambda}$ for each stellar mass on the isochrones.
For each scenario in which two or three objects are involved, we simultaneously fit the observed multicolor light curves and magnitudes.
We take a Bayesian approach with the Markov chain Monte Carlo (MCMC) method 
and estimate the most likely stellar masses on the isochrones and the orbital parameters in $f_{\rm \lambda}$
based on marginalized posterior distributions.
In this process, we employ the variable term in the log-likelihood function $\ln {L_{\rm var}}$ for maximization as follows:
\begin{eqnarray}\label{eq:5}
\ln {L_{\rm var}} ~&=&~ -\frac{1}{2}\chi^2 \nonumber\\
 &=&~ -\frac{1}{2}\sum_{\lambda_{F}}\biggl[ \sum_{i} \frac{(F_{{\rm mod}, \lambda_{F}, i} - F_{{\rm obs}, \lambda_{F}, i})^2}{\sigma_{\lambda_{F}, i }^2}\biggr] \nonumber \\
 &&~~ - \frac{1}{2}\sum_{\lambda_{m}} \frac{({\rm mag}_{{\rm mod}, \lambda_{m}} - {\rm mag}_{{\rm obs}, \lambda_{m}})^2}{\sigma_{\lambda_{m} }^2} ,
\end{eqnarray}

where $\lambda_F$ and $\lambda_m$ indicate the passbands used for the light curve and the apparent magnitude, respectively, 
the subscripts ``${\rm mod}$'' and ``${\rm obs}$" indicate the modeled and observed data, respectively, 
$\sigma_*$ is the uncertainty for each data point, 
and $F$ and $\rm mag$ are the models described above.
We use a quadratic function for the flux baseline in $F$.
In addition, we define $\beta = b \times \frac{R_{\rm cen}}{(R_{\rm cen} + R_{\rm tra})} $, 
where $R$ is the radius of each object and 
the subscripts ``cen" and ``tra" indicate the orbital central star and the transiting/eclipse object, respectively,
 as an alternative to the impact parameter $b$
in order to account for the case that the transiting object is larger than the central star.
In cases where the transiting object is a star (i.e., EB or ET scenarios), 
we can derive all $R$ values 
from the effective temperatures and the luminosities on the isochrones by the Stefan-Boltzmann's law.
On the other hand, for the case of planetary system (i.e., PS or PB scenarios), 
only $R_{\rm tra}$ is independent of the isochrones.

We use the Python package \texttt{emcee} \citep{Foreman-Mackey2013} ensemble sampler for the MCMC in the fitting process and evaluation of uncertainties in the derived quantities.
The fitting parameters in all our models are the time of transit center $t_c$, scaled semi-major axis $a/R_{\rm cen}$,
reduced transit/eclipse impact parameter $\beta$,  quadratic baseline coefficients for the passband $c_{\lambda, 0}$, $c_{\lambda, 1}$ and $c_{\lambda, 2}$, and
central stellar mass $M_{\rm cen}$. %
Additional fitting parameters are the companion star's mass $M_{\rm com}$ for the PB and ET scenarios, radius of the transiting object $R_{\rm tra}$ for the PS and PB scenarios, and 
eclipsing star's mass $M_{\rm tra}$ in the EB and ET scenarios, respectively.
We apply uniform distributions as uninformative priors for these fitting parameters.
Initial positions of the walkers are set to uniform distributions in their domain ranges,
to search the parameter space widely.
There are uncertainties about $\rm mag_{mod,{\lambda}}$ due to the parallax and the extinction vector.
We thus randomly vary these parameters with Gaussian distributions in each step of the chain.
The widths of Gaussians follow the nominal errors in {\it Gaia} parallax measurements and in \citet{Yuan2013} for the extinction. 
After deriving the quantitative physical properties by the MCMC fittings for all scenarios, 
we compare and evaluate each reduced $\chi^2$ (: $\chi^2_{\rm red}$) and Bayesian information criteria \citep[BIC ; ][]{Schwarz1978}, and find the most plausible scenario(s).

\begin{figure*}[t]
 \centering
 \includegraphics[width=18cm]{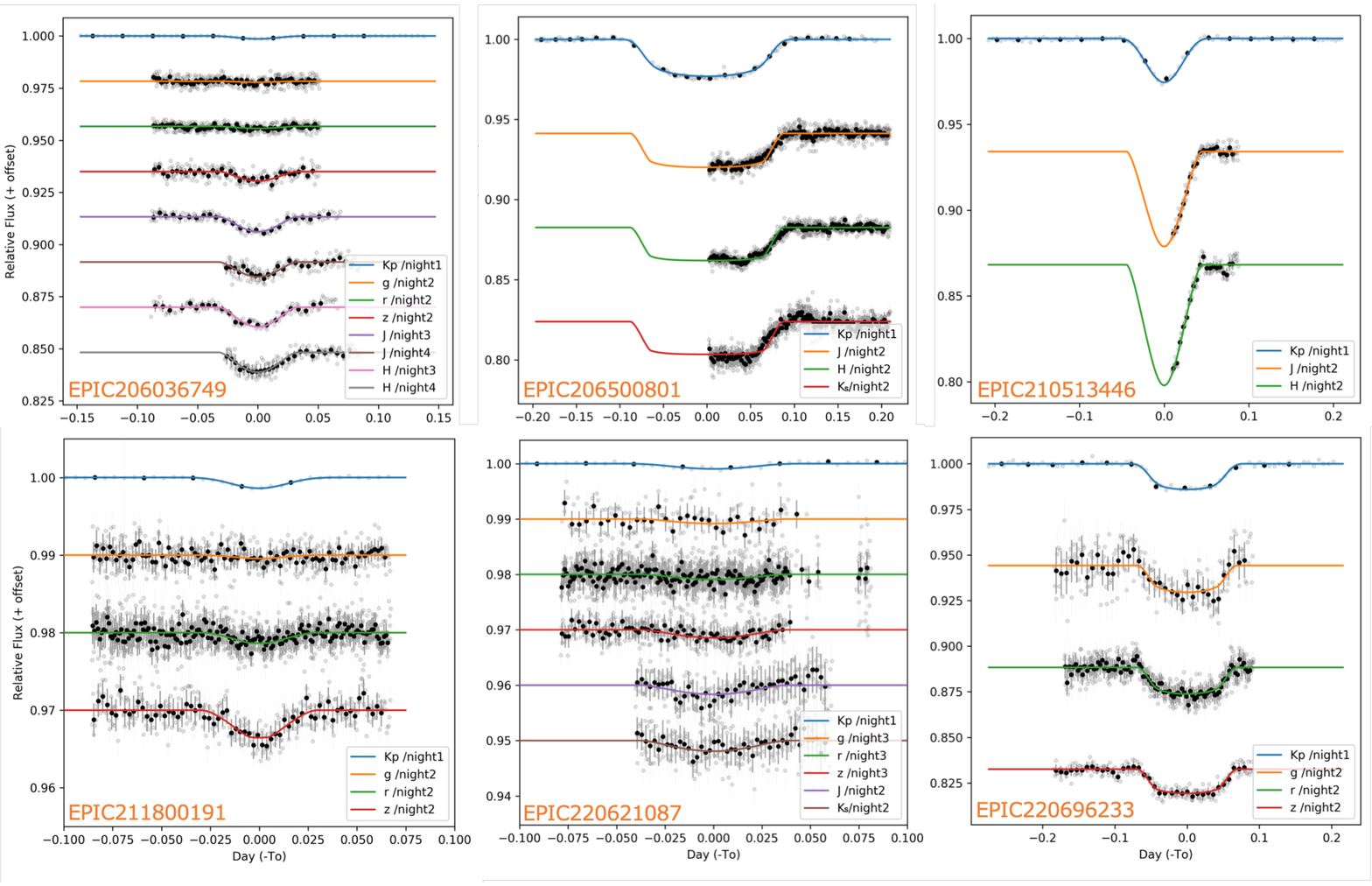}
 \caption{\small Light curves of the targets listed in table \ref{tab:stellar_param} observed by K2, Okayama 188cm/MuSCAT and IRSF/SIRIUS. Observed and binned data point are plotted by gray and black dots, respectively. The best-fit light curves are shown by colored solid lines.}%
\label{lightcurves}
\end{figure*}

\begin{table*}[t]
\centering
\caption{\small Fiducial results with 3.0 Gyr MIST isochrones}\label{tab:res1}
  \begin{tabular}{clcccccccccccc} \hline \hline
    EPIC 		& scenario 	& $M_{\rm tra}(M_{\odot})$ & $R_{\rm tra}(R_{\odot})$& ~	&$M_{\rm cen}(M_{\odot})$ 	& $R_{\rm cen}(R_{\odot})$		&~	
    								&$M_{\rm com}(M_{\odot})$	& $\beta$ & $\chi^2_1$ & $\chi^2_2$ &$\chi^2_{\rm red}$ & BIC\\ \hline
								
    $206036749$	& PS 			& -						& $0.05_{-0.01}^{+0.01}$ 	& & $1.07_{-0.01}^{+0.01}$ 	& $1.03_{-0.01}^{+0.02}$ & &
    								- 						& $0.91_{-0.05}^{+0.01}$	&2479.82 &52.04 & $1.84$	& $2763.90$ \\ 
                 		& PB			& - 						& $0.19_{-0.02}^{+0.02}$ 	& & $0.47_{-0.02}^{+0.04}$	& $0.44_{-0.01}^{+0.03}$& 	&
								$1.10_{-0.01}^{+0.01}$ 	& $0.51_{-0.06}^{+0.06}$	&1714.57 &29.11 & $1.27$	& $1982.97$\\ 
                 		& EB  			& $1.05_{-0.68}^{+0.01}$ 	& $1.00_{-0.63}^{+0.02}$	& & $0.34_{-0.04}^{+0.71} $	& $0.35_{-0.03}^{+0.65}$&	&
								 - 						& $0.54_{-0.14}^{+0.37} $	& 3209.67&48.12 &$2.36$	& $3489.83$\\
                 		& ${\it ET} $ 	& $0.60_{-0.04}^{+0.03}$ 	& $0.57^{+0.04}_{-0.05}$	&  & $0.48_{-0.03}^{+0.03} $ 	& $0.45_{-0.03}^{+0.02}$ & &
								$1.04_{-0.00}^{+0.01}$	& $0.75_{-0.03}^{+0.03} $	& 1666.38& 22.97&$1.23$ 	& $1928.64$\\ \hline
                 
    $206500801 $ & $\it PS$ 	& -		 				& $0.22_{-0.00}^{+0.01}$ 	& & $1.33_{-0.01}^{+0.01} $ 	& $1.57_{-0.03}^{+0.03}$ & &
    								 - 						& $0.21_{-0.01}^{+0.02} $ 	&2953.49& 14.15& $1.45$ 	& $3104.65$ \\ 
                 		& PB  			& - 						& $0.17_{-0.01}^{+0.01}$ 	& & $1.21_{-0.03}^{+0.02} $ 	& $1.28_{-0.06}^{+0.04}$ & &
								 $0.64_{-0.05}^{+0.03}$ 	& $0.12_{-0.03}^{+0.04} $ 	& 3949.84&21.94 &$1.94$ 	& $4116.86$\\ 
                 		& EB  			& $1.22_{-0.13}^{+0.02}$ 	& $1.30_{-0.24}^{+0.05}$ 	& & $1.02_{-0.12}^{+0.03} $ 	& $0.96_{-0.14}^{+0.04}$& &
								-  						& $0.77_{-0.16}^{+0.02} $	& 3916.60&33.38 &$1.92$ 	& $4087.42$\\ 
				& $\rm EB^{*}$			& $0.16^{+0.01}_{-0.01}$ 	& $0.20^{+0.01}_{-0.01}$ &	& $1.25^{+0.03}_{-0.04}$ & 	$1.37_{-0.09}^{+0.07}$& 
								&- 						& $0.42^{+0.05}_{-0.15}$ 	& 2867.08&74.01 &1.43 & 3078.53\\
                 		& ET  			& $0.64_{-0.02}^{+0.03}$ 	& $0.60_{-0.02}^{+0.03}$ 	& & $1.00_{-0.20}^{+0.03} $ 	& $0.93_{-0.20}^{+0.04}$& &
								 $1.16_{-0.28}^{+0.02}$ 	& $0.72_{-0.01}^{+0.02} $	& 3793.31&11.57 &$1.85$ 	& $3949.96$\\ \hline
                 
    $210513446 $ & PS 			& - 						& $0.23_{-0.03}^{+0.94}$ 	& & $0.90_{-0.01}^{+0.02} $ 	& $0.87_{-0.01}^{+0.03}$& &
    								 - 						& $0.72_{-0.04}^{+0.23} $	& 514.90& 22.16&$1.93$ 		& $622.26$ \\ 
                 		& PB  			& - 						& $0.20_{-0.02}^{+0.02}$ 	& & $0.89_{-0.02}^{+0.02} $ 	& $0.86_{-0.03}^{+0.02}$&  &
								$0.33_{-0.10}^{+0.08}$ 	& $0.72_{-0.03}^{+0.02} $ 	&471.78& 22.78&$1.79$ 	& $585.44$\\ 
                 		& $\it EB$  	& $0.82_{-0.01}^{+0.01}$ 	& $0.77_{-0.01}^{+0.02}$ 	& & $0.54_{-0.01}^{+0.01} $ 	& $0.51_{-0.01}^{+0.01}$& &
								 - 						& $0.58_{-0.03}^{+0.03} $ 	& 271.36& 9.58&$1.01$ 	& $366.14$\\
                 		& ET  			& $0.56_{-0.06}^{+0.09}$ 	& $0.53_{-0.06}^{+0.08}$ 	& &$0.48_{-0.03}^{+0.03} $ 	& $0.45_{-0.03}^{+0.03}$& &
								$0.76_{-0.10}^{+0.04}$ 	& $0.43_{-0.05}^{+0.06} $ 	& 272.12&3.39 &$0.99$ 	& $366.39$\\ \hline   
                 
    $211800191 $ & PS 			& - 						& $0.04_{-0.01}^{+0.01}$ 	& & 	$1.05_{-0.01}^{+0.01} $ 	&$1.14_{-0.02}^{+0.02}$& &
    								 - 						& $0.37_{-0.21}^{+0.22} $	&1819.32 & 8.72&$1.11$ 	& $1961.58$ \\ 
                 		& $\it PB$  	& - 						& $0.16_{-0.01}^{+0.01}$ 	& &$0.30_{-0.04}^{+0.06} $ 	&$0.29_{-0.04}^{+0.06}$& &
								 $1.04_{-0.01}^{+0.01}$ 	& $0.14_{-0.09}^{+0.14} $	& 1743.04&6.77 &$1.06$ 	& $1890.77$\\ 
                 		& EB  			& $1.00_{-0.04}^{+0.02}$ 	& $1.04_{-0.08}^{+0.03}$ 	& &$0.57_{-0.03}^{+0.14} $ 	&$0.54_{-0.03}^{+0.12}$& &
								 - 						& $1.00_{-0.07}^{+0.00} $	& 1762.43& 2.63&$1.07$ 	& $1898.60$\\
                 		& ET  			& $0.51_{-0.10}^{+0.10}$ 	& $0.48_{-0.10}^{+0.10}$ 	& &$0.36_{-0.07}^{+0.07} $ 	&$0.34_{-0.06}^{+0.06}$& &
								$1.00_{-0.25}^{+0.02} $ 	& $0.62_{-0.12}^{+0.10} $	& 1744.16&10.73 &$1.06$ 	& $1894.55$\\ \hline
                 
    $220621087 $ & PS 			& - 						& $0.01_{-0.01}^{+0.14}$ 	& &$0.47_{-0.01}^{+0.01} $ 	&$0.44_{-0.01}^{+0.01}$& &
    								 - 						& $0.27_{-0.11}^{+0.86} $ 	&1693.87&112.50&$1.01$ 	& $1993.92$ \\ 
                 		& PB  			& -	 					& $0.01_{-0.01}^{+0.01}$ 	& &$0.36_{-0.03}^{+0.02} $ 	&$0.34_{-0.03}^{+0.02}$& &
								$0.33_{-0.03}^{+0.03} $ 	& $0.31_{-0.18}^{+0.35} $	&1659.44&70.43&$0.97$ 	& $1924.93$\\ 
                 		& EB  			& $0.31_{-0.07}^{+0.05}$ 	& $0.30_{-0.06}^{+0.04}$ 	& &$0.38_{-0.04}^{+0.04} $ 	&$0.36_{-0.04}^{+0.03}$ & &
								-						& $1.10_{-0.12}^{+0.77} $	&1678.33& 70.53&$0.97$ 	& $1936.41$\\
                 		& $\it ET$  	& $0.24_{-0.03}^{+0.03}$ 	& $0.24_{-0.04}^{+0.04}$ 	& &$0.23_{-0.04}^{+0.05} $	&$0.24_{-0.03}^{+0.03}$& &
								$0.33_{-0.05}^{+0.03} $ 	& $0.96_{-0.00}^{+0.01} $	&1657.81&59.13&$0.96$ 	& $1911.19$\\ \hline
                 
    $220696233 $ & PS 			& - 						& $0.07_{-0.01}^{+0.01}$ 	& &$0.61_{-0.01}^{+0.01} $ 	&$0.58_{-0.01}^{+0.01}$& &
    								 - 						& $0.57_{-0.08}^{+0.06} $	&1147.59&21.53&$1.32$ 	& $1291.07$ \\ 
                 		& $\it PB$  	& - 						& $0.06_{-0.00}^{+0.01}$ 	& &$0.54_{-0.01}^{+0.01} $ 	&$0.50_{-0.01}^{+0.01}$& &
								$0.38_{-0.03}^{+0.03} $ 	& $0.51_{-0.11}^{+0.07} $ 	&1146.77&11.18&$1.31$ 	& $1287.30$\\ 
                 		& EB  			& $0.47_{-0.02}^{+0.02}$ 	& $0.44_{-0.02}^{+0.02}$ 	& &$0.47_{-0.02}^{+0.02} $ 	&$0.44_{-0.02}^{+0.02}$& &
								 - 						& $0.88_{-0.01}^{+0.01} $	&1264.74&12.5&$1.44$ 	& $1399.78$\\
                 		& ET 	 		& $0.41_{-0.02}^{+0.03}$ 	& $0.39_{-0.02}^{+0.02}$ 	& &$0.40_{-0.01}^{+0.02} $ 	&$0.38_{-0.01}^{+0.02}$& &
								$0.38_{-0.03}^{+0.02} $ 	& $0.85_{-0.01}^{+0.01} $ 	&1276.78&28.21&$1.47$ 	& $1434.34$\\    
    
    \hline
  \end{tabular}
  \begin{flushleft}
  	{\bf Note :} The minimum BIC scenarios are written in italic. The additional $\rm EB^{*}$ in EPIC 206500801 is the result with adjusting the prior distribution, see Section \ref{ss:00801} for details.
   \end{flushleft}
\end{table*}

\section{Targets and Observations}\label{sec:observation}
\subsection{Targets}
In order to demonstrate our method and software, we selected several candidate transiting systems 
identified by the K2 mission, which is an extended mission of the Kepler prime mission \citep{Borucki2010, Howell2014}.
We successfully observed six systems with multicolor transit photometry 
using the {\bf Mu}lticolor {\bf S}imultaneous {\bf Ca}mera for Studying Atmospheres of {\bf T}ransiting Exoplanets (MuSCAT) mounted on the Okayama 188cm telescope \citep{Narita2015} and/or {\bf S}imultaneous-color {\bf I}nfra-{\bf R}ed {\bf I}mager for {\bf U}nbiased {\bf S}urvey \citep[SIRIUS;][]{Nagashima1999} atop the {\bf I}nfra{\bf r}ed {\bf S}urvey {\bf F}acility (IRSF 1.4 m) telescope at the South Africa Astronomical Observatory.
EPIC ID and stellar parameters of the six targets are summarized in Table \ref{tab:stellar_param} (described in Section \ref{sec:spectro}).
The parallax values were adopted from the {\it Gaia} archive for all targets.
The apparent magnitudes from VizieR are listed in Table \ref{tab:stellar_mgnitudes}.
The orbital period, transit depth, and radius ratio of each candidate in the literature are listed in Table \ref{tab:orbit_param}. 

We downloaded the reduced K2 light curves provided by \citet{Vanderburg2014}. 
We normalized the light curves by fitting it by a fifth-order polynomial function.
We folded the light curves with the orbital periods in Table \ref{tab:orbit_param} 
and removed flux outliers by a $4$ - $\sigma$ clipping for each 0.05 day-long segment.
Finally, we folded the light curves with the orbital periods in Table \ref{tab:orbit_param} and binned into 0.005 day.
Since the six targets listed in Table \ref{tab:stellar_param} cover a wide range of the parameter space 
in terms of the stellar mass, transit depth, and period, etc, 
those targets become a good set of samples to test and demonstrate our methodology 
and software to identify the true nature of the systems.
Below we give a brief summary of each target 
before applying our analysis tool.
\subsubsection{EPIC 206036749}\label{obs:36749}
EPIC 206036749 was identified to be a candidate planet host in K2 Campaign 3
with a $\approx 0.1\,\%$ transit depth and 1.13 day period \citep{Crossfield2016}.
It appears to be a Neptune-sized planetary candidate orbiting a G star based on 
the K2 photometry.
Apparent magnitudes in $V$ and $J$ are 13.2 and 11.7, respectively.
We obtained high contrast AO images of this target with the Infrared Camera and Spectrograph \citep[IRCS; ][]{Tokunaga1998, Kobayashi2000} on the Subaru 8.2 m telescope,
and found that no contaminant sources were resolved in the $20''\times20''$ 
field-of-view (FOV).

\subsubsection{EPIC 206500801}
A Jupiter-sized planet candidate or eclipsing stellar companion was detected around 
EPIC 206500801, a G-dwarf star in Campaign 3 \citep{Barros2016}. 
The orbital period is reported to be 8.16 days. 
This candidate shows a very deep ($\approx 2.5\,\%$), flat-bottomed transit light curve.
Apparent magnitudes in $V$ and $J$ are 12.2 and 10.7, respectively.
No follow-up high-resolution imaging for EPIC 206500801 has been reported.

\subsubsection{EPIC 210513446}\label{obs:13446}
EPIC 210513446 was identified as a host of a Neptune-sized planet candidate 
with a 1.15-day period in Campaign 4 \citep{Crossfield2016}.
The apparent magnitudes in $V$ and $J$ are 14.0 and 11.9, respectively.
Its light curve shows a deep and apparently V-shaped transit signal ($\approx 1.9\,\%$ in the $Kp$ band).
A contaminant star was detected by AO observations with the Gemini-N 8-m telescope / NIRI 
\citep{Klaus2003} with $\Delta {\rm mag} \approx 1$ at $\approx 0^{''}.5$ away
from the central star, according to the ExoFOP-K2\footnote{https://exofop.ipac.caltech.edu/k2/}
website.

\subsubsection{EPIC 211800191}\label{obs:00191}
EPIC 211800191 is reported to be a candidate planetary system in Campaign 5 \citep{Barros2016}.
The transit depth and the orbital period are $\approx 0.1\,\%$ and 1.11 days,
corresponding to a Neptune-sized planet around a G star.
The properties of the system are very similar to those of EPIC 206036749.
The apparent magnitudes in $V$ and $J$ are 12.6 and 11.4, respectively.
AO images by Keck II/ NIRC2 \citep{Wizinowch2000} can be found at
the ExoFOP archive, but no contaminant source can be found
in the $6''\times6''$ FOV.

\subsubsection{EPIC 220621087}\label{obs:21087}
EPIC 220621087 (K2-151) was already validated as a planetary system, 
hosting an Earth-sized planet around a $0.44\,M_{\odot}$ M dwarf in Campaign 8 \citep{Hirano2018a}.
The transit depth and the orbital period in the K2 light curve are $\sim 0.1\,\%$ and 3.8 days, respectively. 
The apparent magnitudes in $V$ and $J$ are 14.1 and 10.9, respectively.
While the planetary nature of the target was confirmed before, we reanalyze 
the multicolor photometric data for this target to demonstrate our tool.

\subsubsection{EPIC 220696233}\label{obs:96233}
 EPIC 220696233 is an early M dwarf, hosting a Jupiter-sized planetary candidate 
 in Campaign 8 \citep{Livingston2018, Dressing2019}.
 The transit depth and the orbital period are $\sim 1.2\,\%$ and 28.73 days, 
 respectively. 
 The apparent magnitudes in $V$ and $J$ are 16.2 and 13.2, respectively.
 A high-contrast image is available on the ExoFOP archive, taken by the
 Palomar 5 m telescope \citep{Hinkley2008,Hinkley2011}, but no companion 
 was found in the $8''\times8''$ FOV.
 This target is flagged as a possible eclipsing binary.

\subsection{Spectroscopic Parameters}\label{sec:spectro}
 To better characterize our targets, we determined the stellar properties of the targets 
 using the archived data available.
 For EPIC 206036749, EPIC 210513446, EPIC 211800191 and EPIC 220621087, 
 we downloaded the high-resolution spectra observed by Keck/HIRES available at ExoFOP-K2.
 For EPIC 206500801, a spectrum observed with HARPS-N \citep{Cosentino2012} was downloaded 
 from IA2/TNG archived website\footnote{http://archives.ia2.inaf.it/tng/}.
We made use of \texttt{SpecMatch-Emp} \citep{Yee2017} to derive stellar properties 
($T_{\rm eff}$, $R_*$, $M_*$ and [Fe/H]) of the targets from high-resolution spectra. 
These high-resolution spectra were analyzed following \citet{Hirano2018a}.
 For EPIC 220696233, for which no high-resolution spectrum was available,
 we adopted the values of effective temperature, metallicity, and $\log{g}$ 
 listed on ExoFOP-K2 \citep{Huber2016}, but we derived 
its stellar radius and mass using the empirical relations for low-mass stars \citep{Mann2015}
based on the above parameters and Gaia parallax.
The results of these analyzes are summarized in Table \ref{tab:stellar_param}.
Those parameters are helpful to understand the properties of the primary (brightest) star for each system, but 
in the following analyses, we only use the derived metallicities in generating the 
isochrones during the fitting process, as there are large uncertainties in spectroscopically derived parameters when the system includes multiple stars.
The other stellar parameters in Table \ref{tab:stellar_param} 
are used for comparisons of the derived results.

\subsection{Observations}

\subsubsection{\it OAO, ~188cm / MuSCAT}
 We conducted follow-up transit/eclipse photometry for EPIC 206036749, EPIC 220621087, EPIC 211800191, and EPIC 220696233 using MuSCAT \citep[][]{Narita2015}
 on UT 2015 August 24, 2016 September 20, 2016 November 25, and  2017 February 15, respectively. 
 We simultaneously obtained three-band images with MuSCAT, consisting of three $1k \times 1k$ CCDs
 with the pixel scale of $0\farcs36$ pixel$^{-1}$ through the SDSS 2nd-generation $g'$, $r'$, and $z_s$-band filters.
 
  The exposure times were set to 30 and 60 s for EPIC 206036749, 25, 12 and 35 s for EPIC 211800191, 60, 10 and 25 s for EPIC 220621087 and 120, 60 and 120 s for EPIC 220696233 in $g'$, $r'$ and $z_s$, respectively.
  We completed the observations with photometric conditions, 
  covering the whole transits/eclipse except EPIC 220621087.
  The observation of EPIC 220621087 was interrupted for $\sim 0.9$ hour
  due to cloud passage after the transit egress.

\subsubsection{\it SAAO,~IRSF / SIRIUS}
 Between UT 2016 September 25 and October 10, we conducted multicolor transit photometry 
 for EPIC 206036749, EPIC 206500801, EPIC 210513446, EPIC 220621087 using SIRIUS \citep[][]{Nagashima1999,Nagayama2003}. 
 SIRIUS is equipped with three 1k $\times$ 1k HgCdTe detectors with pixel scale of $0\farcs45\,{\rm pixel}^{-1}$. 
 It enables us to take three near-infrared images in the $J$, $H$ and $K_s$ bands simultaneously 
 with the same exposure time for all bands.
 
 EPIC 206036749 and EPIC 210513446 are faint ($\approx11$ in $J$), and in order to 
 achieve sufficiently high signal-to-noise (S/N) ratio of light curves, we set the integration times to $50 - 100$ seconds.
 This resulted in the $K_s$-band images being saturated, 
 and thus we discarded the $K_s$-band data from the analysis.
 Regarding EPIC 206036749, we successfully observed two transits.
 We observed EPIC 206500801 and EPIC 220621087 with the exposure time of 30 s.
 However, we found that the brighter comparison star of EPIC 220621087 
 was saturated in the $H$ band by mistake.
 The images of EPIC 206500801 and EPIC 210513446 were obtained 
 only after the transit center, due to targets' visibilities and telescope technical trouble.
 
 Aperture photometry for each target was performed with a customized pipeline described in \citet{Fukui2011}.
 The resulting light curves are shown in Figure \ref{lightcurves}.
 For EPIC 206036749, EPIC 210513446, and EPIC 211800191, 
 the transit/eclipse depths become deeper at longer wavelengths.
 For the other systems, on the other hand, 
 the depth variations were found to be relatively small by visual inspections.
 \begin{figure*}[t]
 \centering
 \includegraphics[width=18cm]{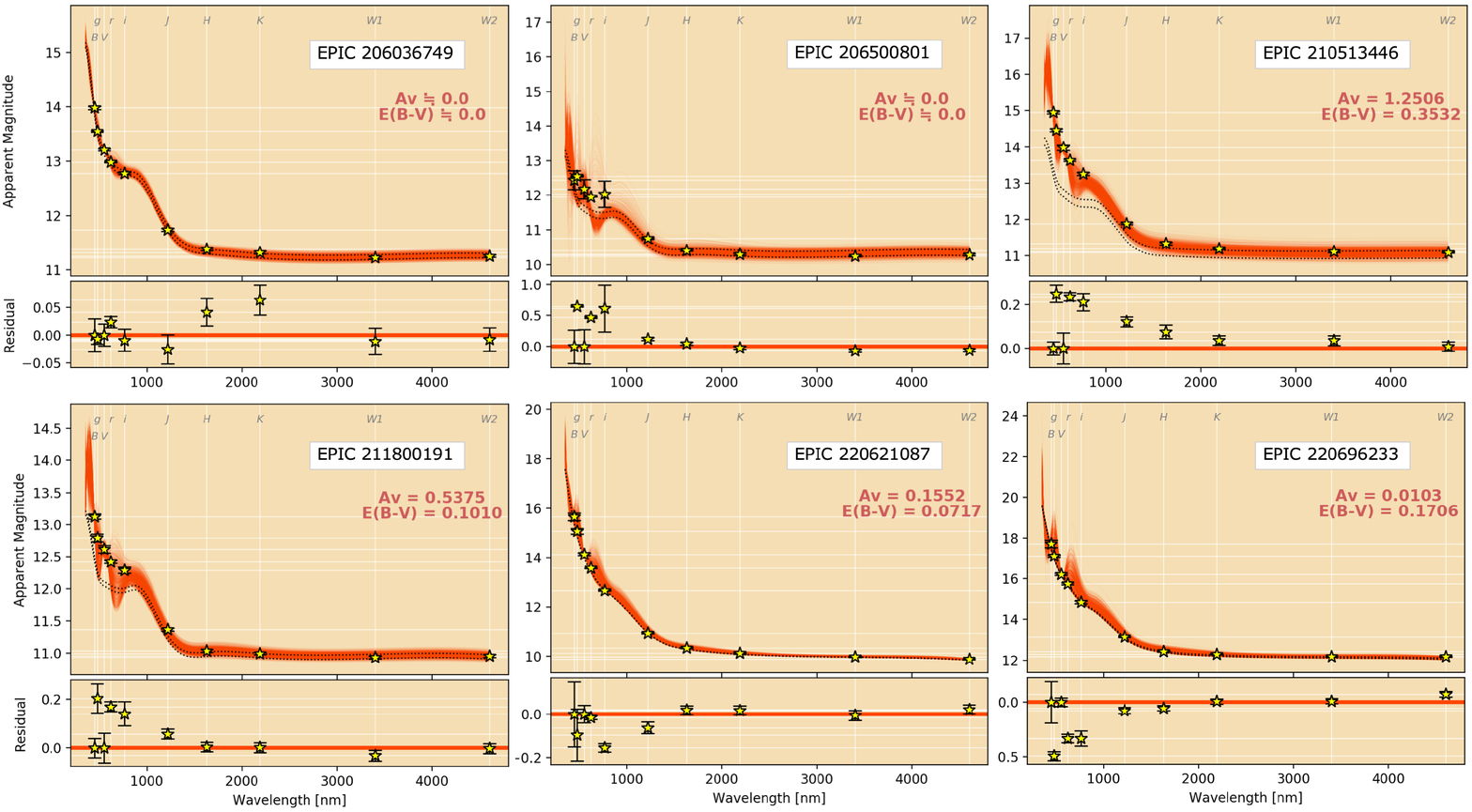}
 \caption{\small The first result of SED term derived in Section \ref{sec:results}. Top panel for each target : Horizontal axis represents the median wavelength of each passband and vertical axis represents apparent magnitude. The star points are observed values listed in Table \ref{tab:stellar_mgnitudes}. 
 The width of the red blurred line represents the Gaussian uncertainties of the $\it Gaia$ parallax, $E(B-V)$, and the extinction vector.
 The black dashed lines represent the models that do not include reddening and have widths corresponding to the $\pm 1\sigma$ uncertainty of the parallaxes. We list the extinction module in $V$ band, $Av$ and the $E(B - V)$ for each target. Bottom panel for each target : Residual plots from the mean of the models.} %
\label{photometry}
\end{figure*}

\begin{figure*}[t]
 \centering
 \includegraphics[width=15cm]{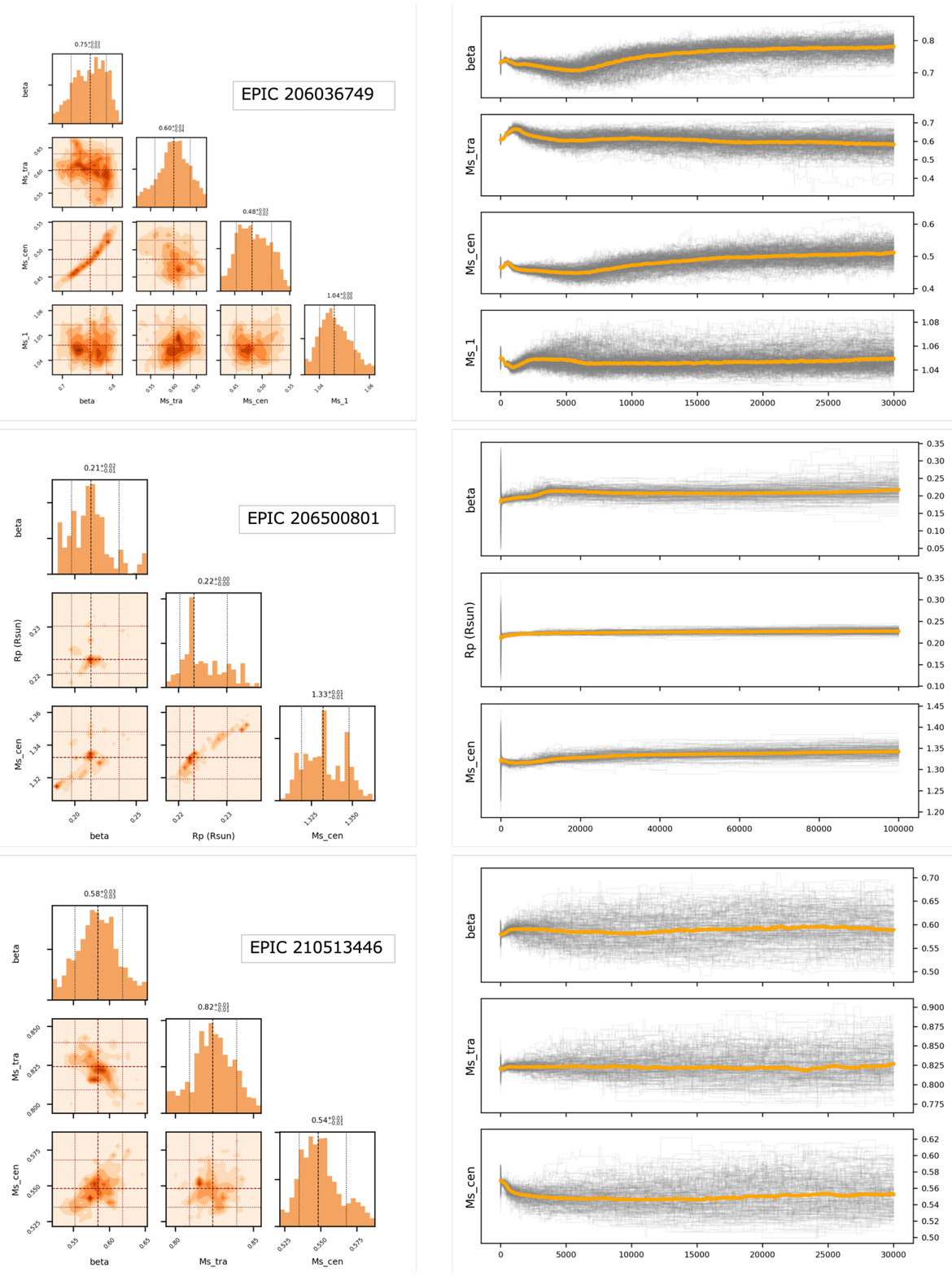}\\
 \caption{\small Left panel of each target : Corner plot of the posterior distribution for each target with the most likely scenario in Table \ref{tab:res1}, where $beta$ is reduced impact parameter; $Ms_{\rm tra}$, $Ms_{\rm cen}$ and $Ms_1$ are eclipsing star mass, host star mass, and contaminated star mass, respectively; $Rp$ is radius of the planetary candidate.
 We set the ranges of the plot within $\pm 2 \sigma$. 
 Right panel of each target : Plot of the walker with the MCMC steps. We use the last $10^4$ steps for the marginalization of the left panels. Orange line represents medians of the total walkers. }%
\label{corner}
\end{figure*}

\begin{figure*}[t]
 \centering
 \includegraphics[width=15cm]{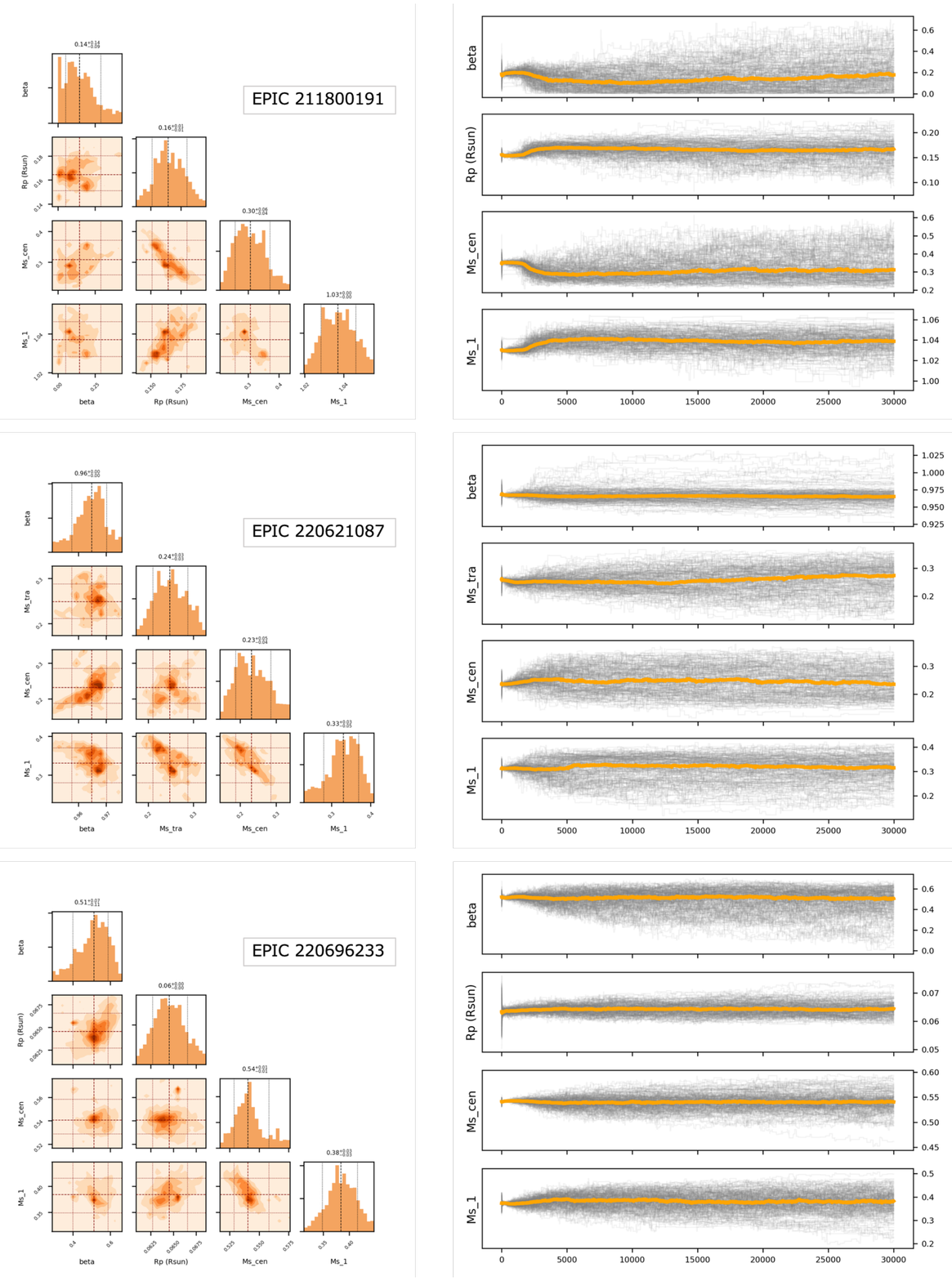}\\
 \caption{\small Continued plot of Figure \ref{corner}.}%
\label{corner2}
\end{figure*}

\begin{figure*}[t]
 \centering
 \includegraphics[width=18cm]{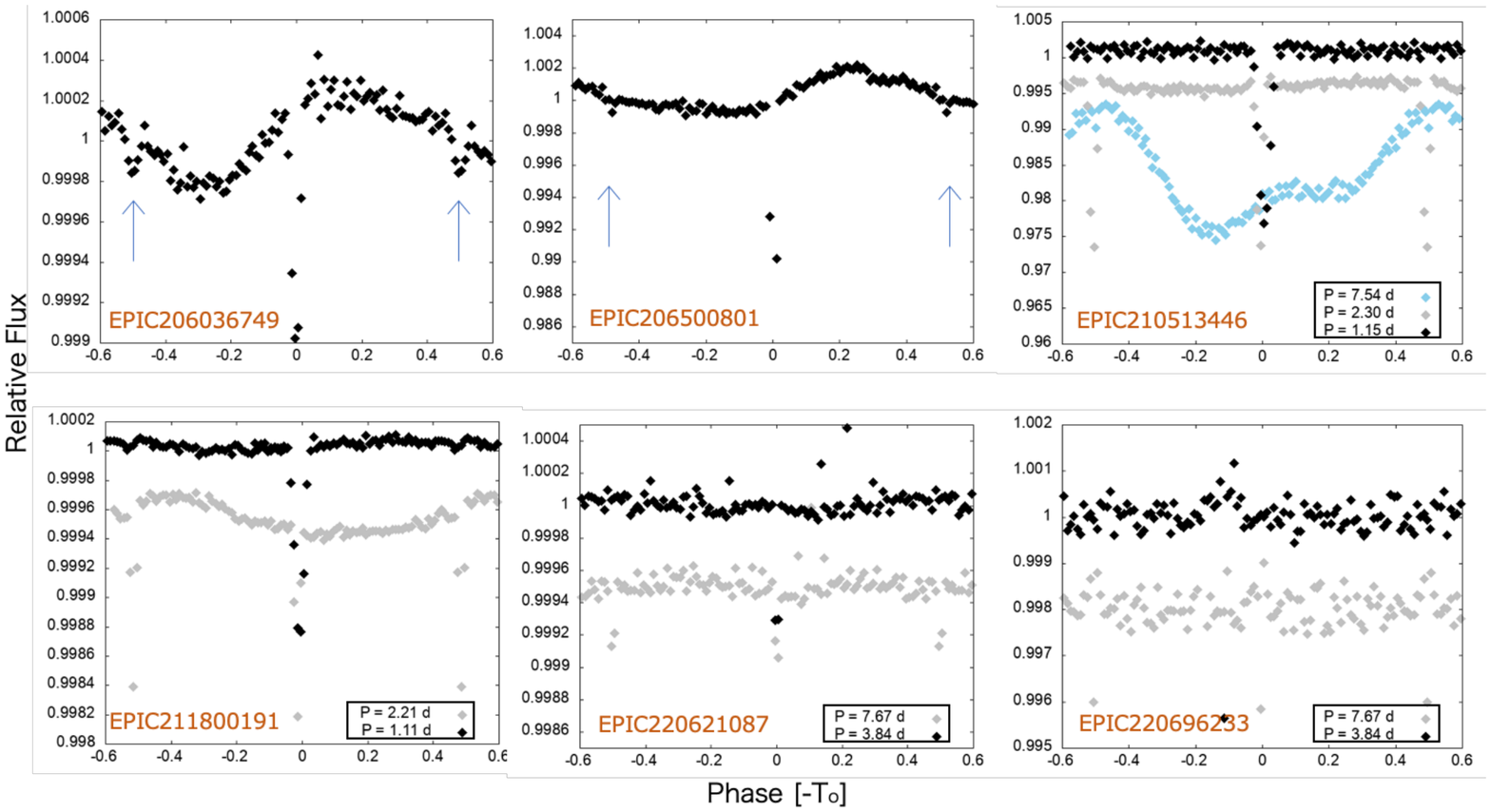}
 \caption{\small The phase curves with 0.01-sized bin for all the targets. The light curves with black dots are folded in the period in Table \ref{tab:orbit_param}. The shallow secondary eclipse-like signals are indicated by the arrow for EPIC 206036749 and EPIC 206500801. The grey dots are folded in the double orbital period to check the secondary eclipse.  The blue dots in EPIC 210513446 are detected in a period of 7.54 days, and are considered to be due to the companion star.}%
\label{reflc}
\end{figure*}

\section{Analyses and Results}\label{sec:results}
We applied our algorithm described in Section \ref{sec:method} to the observed light curves \citep{Vanderburg2014} and apparent magnitudes
 for each of the six targets to retrieve the physical properties under our scenarios.
We set the age of isochrones to 3.0 Gyr as a representative value,
because the ages for our targets are not known.
The metallicity for the isochrones is set to the value listed in Table \ref{tab:stellar_param}.
We will discuss the impact of the assumed age and metallicity assumed for the isochrones in Section \ref{sec:discussion}.

We used a quadratic law for the limb darkening of the transit model, 
and its coefficients were held fixed at theoretical values \citep{Claret2011}\footnote{http://vizier.cfa.harvard.edu/viz-bin/VizieR}
based on each primary star's atmospheric parameters (Table \ref{tab:stellar_param}). 
This is because most of our targets are relatively faint and the photometric precisions 
achieved by our ground-based observations are not sufficiently high to properly evaluate them.
Considering the computational efficiency, 
we set the orbital periods to the values in Table \ref{tab:orbit_param}.
Similarly, we fixed orbital eccentricity at $e=0$ . 
For the MCMC ensemble sampler \citep{Foreman-Mackey2013}, 
we set the number of walkers to 200, except EPIC 206036749.
Since EPIC 206036749 has a large number of fitting parameters to correspond with the number of the observed light curves, 
we set the number of walkers to 300.
After we ran the sampler for 30,000 steps to find the optimal parameter space,
we performed additional 30,000 steps by setting the initial positions of walkers around the optimized values.
For EPIC 206500801, we retried the sampling for 100,000 steps, because the convergence was not clear.
Finally, we ensured that all the parameters converge visually in half of the steps for all the targets,
and confirmed that the chain lengths are sufficiently long enough for the analyses and discussion.

The derived parameters in each scenario are listed in Table \ref{tab:res1} for the 
fiducial case (i.e., the MIST isochrones, 3 Gyr); 
$M_{\rm tra}$, $M_{\rm cen}$, and $M_{\rm com}$ represent the masses of 
transiting/eclipsing object, central (transited/eclipsed) star, and companion star, respectively.
The radius of the transiting/eclipsing object is expressed by $R_{\rm tra}$.
In EB and ET scenarios, $R_{\rm tra}$ is derived from the isochrones using the Stefan-Boltzman's law.
The uncertainties are calculated based on the 68.3 \% range from the median of the 
marginalized posterior distribution.
We show the scenario with the minimum BIC (most likely scenario) for each target in italics
in Table \ref{tab:res1}.
We list the $\chi^2$ values for  the light curve and magnitude models as $\chi^2_1$ and $\chi^2_2$, respectively.
The best-fit light curves and SEDs for the minimum-BIC scenario are shown in Figures \ref{lightcurves} and \ref{photometry}, respectively.
The posterior distributions and behaviors of walkers of each target in the most likely scenarios is shown in Figure \ref{corner} and \ref{corner2}.
We marginalized the final $10^4$ steps for the corner plot and the estimations of the uncertainties.
We also show photometric phase curves with 0.01-sized bins for all targets in Figure \ref{reflc} to further interpret and discuss our results.

 In the above analysis, we adopted the MIST isochrones \citep{Dotter2016, Choi2016} 
 in deriving the parameters of possible unresolved stars. 
 However, isochrone models are generally known to have systematic errors in the derived parameters, 
 and those errors could be particularly significant for very low-mass stars, 
 for which stellar parameters are not well-calibrated by observations. 
 In order to evaluate the impact of those systematic errors arising from adopting specific isochrone models,
 we repeated the analyses above using two different isochrones:
 the Dartmouth Stellar Evolution Program \citep[DSEP;][]{Dotter2008}
 and the stellar tracks and isochrones 
 with the PAdova and TRieste Stellar Evolution Code \citep[PARSEC;][]{Bressan2012}.
 In the analyses, the age was set to 3.0 Gyr as in the case of the MIST isochrones.
 Table \ref{tab:res_isochrones} summarizes the resulting scenarios 
 with the minimum BIC values for the each target for different isochrones.
 Below, we describe the interpretation of the fitting result for each target.

\begin{table*}[t]
\centering
\caption{\small The results of minimum BIC scenarios with DSEP and PARSEC isochrones for each target.}\label{tab:res_isochrones}
  \begin{tabular}{ccccccccccccc} \hline \hline
    EPIC 				& Isochrones & scenario 	& $M_{\rm tra}(M_{\odot})$ 	& $R_{\rm tra}(R_{\odot})$&	~&$M_{\rm cen}(M_{\odot})$ 	&$R_{\rm cen}(R_{\odot})$& ~&
    					$M_{\rm com}(M_{\odot})$	& $\beta$ & $\chi^2_{\rm red}$ & BIC\\ \hline
    $206036749 $ 	&DSEP 		& ET 	& $0.64_{-0.07}^{+0.40}$	& $0.60_{-0.06}^{+0.39}$ 	& & $0.45_{-0.19}^{+0.06}$ 	&$0.43_{-0.15}^{+0.05}$& &
    					$1.02_{-0.37}^{+0.03}$&$0.72_{-0.25}^{+0.06}$&$1.36$&$2116.51$	\\
					&PARSEC 	& ET 	& $0.58^{+0.08}_{-0.07}$ 	& $0.58^{+0.07}_{-0.07}$ 	& & $0.49^{+0.06}_{-0.04}$ 	&$0.48_{-0.04}^{+0.06}$& &
					$1.04^{+0.02}_{-0.29}$ 	& $0.71^{+0.06}_{-0.05}$ 	& $1.23$	& $1938.72 $\\
                  \hline
                 
    $206500801 $ 	& DSEP		& PS 	& - 					& $0.18^{+0.01}_{-0.02}$ 	& & $1.22^{+0.02}_{-0.05}$ 	&$1.32_{-0.10}^{+0.05}$& &
    					- 					& $0.36^{+0.03}_{-0.03}$ 	& $1.42$	& $3063.09$ \\ 
                			&PARSEC 	& PB  	& - 					& $0.18^{+0.02}_{-0.03}$ 	& & $1.09^{+0.04}_{-0.07}$ 	&$1.07_{-0.09}^{+0.07}$& &
					$0.93^{+0.06}_{-0.12}$ 	& $0.50^{+0.03}_{-0.03}$	& $1.37$	& $2960.69$\\ 
                  \hline
                 
    $210513446 $ 	& DSEP		& ET 	& $0.55_{-0.05}^{+0.03}$ 	&  $0.53_{-0.05}^{+0.03}$	& & $0.50_{-0.03}^{+0.03} $ 	&$0.48_{-0.03}^{+0.03}$& &
    					$0.77_{-0.03}^{+0.03}$ 	& $0.45_{-0.06}^{+0.07} $	& $1.00$ 	& $366.72$ \\ 
                			&PARSEC 	& ET  	& $0.58_{-0.06}^{+0.07}$ 	& $0.56_{-0.06}^{+0.06}$ 	& & $0.48_{-0.03}^{+0.04} $ 	&$0.46_{-0.03}^{+0.04}$& &
					$0.74_{-0.08}^{+0.04} $ 	& $0.41_{-0.05}^{+0.06} $ 	& $0.98$ 	& $362.77$\\ 
                  \hline   
                 
    $211800191 $ 	&DSEP		& ET 	& $0.58_{-0.07}^{+0.07}$ 	& $0.56_{-0.05}^{+0.07}$ 	& & $0.45_{-0.04}^{+0.04} $ 	&$0.43_{-0.04}^{+0.04}$& &
    					$0.99_{-0.03}^{+0.02} $ 	& $0.75_{-0.06}^{+0.05} $	& $1.06$ 	& $1891.49$ \\ 
                			& PARSEC	& ET 	& $0.57_{-0.09}^{+0.10}$ 	& $0.55_{-0.09}^{+0.07}$ 	& & $0.43_{-0.05}^{+0.04} $ 	&$0.41_{-0.05}^{+0.04}$& &
					$0.99_{-0.04}^{+0.02} $ 	& $0.71_{-0.10}^{+0.07} $	& $1.07$ 	& $1897.78$\\ 
                  \hline
                 
    $220621087 $ 	&DSEP		& ET 	& $0.30_{-0.04}^{+0.05}$ 	& $0.29_{-0.03}^{+0.32}$ 	& & $0.25_{-0.06}^{+0.06} $ 	&$0.25_{-0.04}^{+0.05}$& &
     					$0.28_{-0.03}^{+0.06} $	& $0.96_{-0.01}^{+0.01} $ 	& $0.93$ 	& $1864.78$ \\ 
                			& PARSEC	& PB  	& - 						& $0.02_{-0.01}^{+0.01}$ 	& & $0.20_{-0.03}^{+0.05} $ 	&$0.22_{-0.03}^{+0.04}$& &
					$0.44_{-0.02}^{+0.01} $ 	& $0.63_{-0.36}^{+0.21} $	& $0.94$ 	& $1872.26$\\ 
                 \hline
                 
    $220696233 $ 	&DSEP		& PS 	& - 						& $0.07_{-0.01}^{+0.01}$ 	&& $0.61_{-0.01}^{+0.01} $ 	&$0.58_{-0.01}^{+0.01}$&&
    					 - 						& $0.67_{-0.06}^{+0.07} $	& $1.31$ 	& $1281.32$ \\ 
                			& PARSEC	& PS  	& -	 					& $0.07_{-0.01}^{+0.01}$ 	& &$0.60_{-0.01}^{+0.01} $ 	&$0.59_{-0.01}^{+0.01}$& &
					 - 						& $0.57_{-0.10}^{+0.06} $ 	& $1.31$ 	& $1284.29$\\ 
    \hline
  \end{tabular}
\end{table*}

\subsection{EPIC 206036749}\label{ss:36749}
The most likely scenario for EPIC 206036749 is ET. 
There are good agreements between the observed quantities and their models for both light curves and magnitudes (Figures \ref{lightcurves} and \ref{photometry}).
In this scenario, the system consists of an eclipsing star ($0.60 \,M_{\odot}$), 
the central star ($0.48\,M_{\odot}$), and the companion star whose mass is  $1.04\,M_{\odot}$.
The eclipsing star is estimated to be larger than the central star, 
suggesting that the reported transit-like events are actually secondary eclipses.
The second possible scenario, although much less likely according to the large $\Delta \mathrm{BIC}$, is PB consisting of a star-sized ($0.19\,R_{\odot}$) dark object,
a central star with $0.47\,M_{\odot}$, and a companion star with $1.10\,M_{\odot}$. 
The other two scenarios (PS and EB) are very unlikely, since both $\chi^2_{\rm red}$ and BIC values are much worse than the above two scenarios.

In both of the results for DSEP and PARSEC, the ET scenario is selected 
as the minimum BIC model as shown in Table \ref{tab:res_isochrones}.
The derived properties are consistent within $\approx 1\sigma$ among the three 
different isochrone models.
The uncertainties for the eclipsing star's mass and the companion star's mass in DSEP are large due to the bimodal posterior distributions of the fitting parameters.
The highest peak is compatible with the results for the other two isochrones,
and the second peak is located around $1.0\,M_{\odot}$ - $0.4\,M_{\odot}$ - $0.6\,M_{\odot}$ 
for $M_{\rm tra}$ - $M_{\rm cen}$ - $M_{\rm com}$, respectively.
This disagreement may be due to the fact that the combination of these three stellar masses degenerates with respect to the SED, although the reason why the bimodal distribution was only seen for DSEP is unclear.

Figure \ref{reflc} indicates that the secondary eclipse (phase near 0.5) is much shallower than the primary transit in the K2 light curve.
This suggests either that the orbital inclination and eccentricity are large (grazing eclipses) or that the transiting/eclipsing object is much cooler than the eclipsed star.
The former case is possible in the ET scenario, 
but the possibility of a large eccentricity has to be low due to the short orbital period (1.13 days)
and the observed timing of the secondary eclipse near phase $\approx 0.5$. 
On the other hand, the latter possibility is compatible with the PB scenario
in Table \ref{tab:res_isochrones}; in the PB scenario, the transiting object is
assumed to be ``dark", meaning that it does not affect the color-dependence of the
transit light curves and SED, but this scenario also allows for a very late-M dwarf or a brown dwarf as the transiting object, which would give negligible impacts on the observable quantities. The detection of the secondary eclipse is indicative of the possibility that the transiting/eclipsing object is actually a very low-mass star rather than a giant planet. 

The folded light curve also exhibits a phase-shifted sinusoidal modulation.
This modulation is unlikely to be caused by ellipsoidal variations,
in which case it would show two peaks in one orbit \citep{Morris1993, Mazeh2010}.
The flux peak/bottom phase is not consistent with the Doppler boosting, either. 
It may be ascribed to interactions of close-in binary stars such as the O'Connell effect \citep{Wilsey2009}. 
As we noted in Section \ref{obs:36749}, 
the contamination star is not resolved within $0\farcs5$, corresponding to $\sim 200$ au in the AO image.
This is a tight constraint for a triple-star systems.
In \citet{Crossfield2016}, EPIC 206036749 is reported to have a Neptune-sized planetary candidate, but our analysis implies a compact multiple-star system (i.e., ET or PB).

\subsection{EPIC 206500801}\label{ss:00801}
The most likely scenario for EPIC 206500801 is PS, consisting of a $0.22\,R_{\odot}$ transiting/eclipsing object and the central star with $1.33\,M_{\odot}$.
The $\chi_{\rm red}^2$ and BIC in the other scenarios are much worse than those for PS.
The minimum BIC scenario with DSEP is also PS, in which the derived parameters in Table \ref{tab:res_isochrones}
are consistent within $\approx 1\sigma$ with those of MIST.
On the other hand, the analysis with PARSEC favors PB, consisting of a $0.18\,R_{\odot}$ transiting object, central star with $1.09\,M_{\odot}$, and companion with $0.93\, M_{\odot}$.
This disagreement in the scenario selections may be due to the difference in SED models for different isochrones.

When we naively interpret the observed transit/eclipse depths for EPIC 206500801,
an EB scenario with e.g, $\approx 0.2\,M_{\odot}$ - $\approx 1.3\, M_{\odot}$ for $M_{\rm tra}$ - $M_{\rm cen}$ is also expected as a plausible solution, 
but this possibility was ruled out in the above analysis.
To further investigate this, we reanalyzed the same data set 
imposing a prior for the eclipsing star being less massive than $0.2\, M_{\odot}$.
In this case, the solution for the EB scenario was found to be $0.16\, M_{\odot}$ - $1.25 \, M_{\odot}$ for $M_{\rm tra}$ - $M_{\rm cen}$, where the $\chi^2_{\rm red}$ and BIC values were 1.43 and 3078.53, respectively, 
which are also listed in Table \ref{tab:res1} as $\rm EB^*$ in addition to the other scenarios.
While the BIC value is the lowest than those for all the scenarios without a prior, the goodness of fit in $\chi_2^2$ ($\chi^2$ for the apparent magnitudes) is the worst among all scenarios. 
The disagreement between the fitting results with and without the prior suggests 
that there are multiple local mimima in $\chi^2$, leading to a degeneracy among
the fitting parameters. The MCMC analysis without a prior explored a wider range
of the parameter space, but this degeneracy of the fitting parameters prohibited
a robust estimation.


Consequently, we were unable to conclude if EPIC 206500801 is a PS or EB.
This is because the flux from a low-mass star less massive than $0.2\, M_{\odot}$ is negligible compared to the solar-type stars in the total SED, and we cannot distinguish between a giant transiting planet and a very low-mass star. 
Fortunately, it is relatively straightforward to identify such a close-in binary, by 
e.g., making a few radial-velocity measurements. 
Note that binary systems whose mass ratio is lower than 0.2 
and orbital period is shorter than 10 days are not so common \citep[see Figure 17 in][]{Raghavan2010}.
In Figure \ref{reflc}, the phase curve looks sinusoidal,
but its exact mechanism is not known as in the case of EPIC 206036749.
A weak secondary eclipse-like signal ($\sim 0.1 \%$) can be seen with the blue arrow in Figure \ref{reflc}, making EB a more likely scenario for this system. 
Further follow-up observations would be required to make a definitive conclusion. 
%

\subsection{EPIC 210513446}
The best-fit scenarios for EPIC 210513446 are EB and ET, which 
are equally likely based on the BIC values in our analysis.
In the EB scenario, masses of the eclipsing star and the central star 
are $0.82\,M_{\odot}$ and $0.54\,M_{\odot}$, respectively, 
and the observed eclipse is actually a secondary eclipse.
In the ET scenario, the system consists of an eclipsing star with $0.56\,M_{\odot}$,
the central star with $0.48\,M_{\odot}$, and the companion star with $0.76 \,M_{\odot}$.
For both DSEP and PARSEC models, 
the ET scenario was selected as the minimum BIC model as in Table \ref{tab:res_isochrones}.
The derived physical parameters are consistent within $\approx 1\sigma$ with those for MIST.

As we noted in the Section \ref{obs:13446}, a contaminant source was resolved in $\approx 0''.5 $, 
corresponding to $\sim$ 160 au with $\Delta {\rm mag_{K}} \approx  1$ in the AO image.
This magnitude difference is almost consistent with the difference between
a $0.6\,M_{\odot}$ - $0.5\,M_{\odot}$ binary and a $0.8\,M_{\odot}$ companion star in our ET scenario.
If the system includes similar-mass eclipsing binary stars, we would observe
secondary eclipses whose depths are similar to the primary, but
the secondary eclipse as well as the photometric variations synchronized with the eclipses was not observed with the 1.15-day period in Figure \ref{reflc} (black dots). 
We did not find phase variations in the light curve folded with the double period (2.30 days; grey dots in Figure \ref{reflc}), either.
On the other hand, we detected a large periodic variation at 7.54 days 
based on the generalized Lomb-Scargle periodogram \citep{Zechmeister2009},
which is shown with blue dots in Figure \ref{reflc}.
Given the large amplitude of this flux modulation, this periodicity at 7.54 days most likely comes from the companion star, emitting the largest flux among the three stars in the system. 
Since the 2.30-day orbital period more consistently explains the observed data (i.e., the secondary eclipse), 
we conclude that EPIC 210513446 is likely to be an ET including an eclipsing binary ($0.6\,M_{\odot}$ - $0.5\,M_{\odot}$) with a period of 2.30 days.


\subsection{EPIC 211800191}
The minimum BIC scenario is PB consisting of a $0.16\, R_{\odot}$ transiting object, 
the central star with $0.30\,M_{\odot}$ and the companion star with $1.04\,M_{\odot}$,
but derived values of $\chi^2_{\rm red} $ and BIC are similar between PB and ET scenarios, and we cannot select the most likely scenario from this analysis.
The ET scenario was selected as the minimum BIC model (Table \ref{tab:res_isochrones})
for both DSEP and PARSEC models.
The derived values of the physical parameters are consistent within $\approx 2\sigma$ with those of MIST.

For the EB and ET scenarios, secondary eclipses would be seen in the K2 light curve, 
but were not detected.
Thus, we suspected that the orbital period in Table \ref{tab:orbit_param} was incorrectly estimated.  
We folded the K2 light curve with the double period,
and found the synchronized photometric phase curve and two eclipses as shown by grey dots in Figure \ref{reflc}.
We checked the ephemeris and reanalyzed the K2 light curve using the same data set and conditions as above, but with the double period.
As a result, we derived the ET scenario as the most plausible one with $0.50 - 0.39 - 1.01\,M_{\odot}$ for the MIST isochrone, which is almost the same as the original result in Table \ref{tab:res1} within $1\,\sigma$.
If the EB or PS scenarios were true, then the secondary eclipse would be much shallower than the primary transit, which was not observed. 
We thus conclude that ET is the most reasonable scenario for the EPIC 211800191 system.
As we noted in Section \ref{obs:00191}, no apparent contaminant was resolved within $0\farcs5$, corresponding to $\sim 200$ au in the AO image.

%
\subsection{EPIC 220621087}\label{ss:21087}
The minimum BIC scenario for MIST is ET, consisting of the eclipsing star with $0.24\,M_{\odot}$, 
the central star with $0.23\,M_{\odot}$, and the companion star with $0.33\,M_{\odot}$.
The result for DSEP selects the ET scenario as the minimum BIC model as in Table \ref{tab:res_isochrones},
which is consistent within $\approx 1\sigma$ with that for MIST.
On the other hand, the result with PARSEC favors PB scenario, 
consisting of a planet with a size of $0.02\,R_{\odot}$,
the central star with $0.20 \,M_{\odot}$, and the companion star with $0.44 \,M_{\odot}$.

As we noted, EPIC 220621087 was already validated as a true planet host \citep[K2-151:][]{Hirano2018a}.
In our assessment, however, EPIC 220621087 is a possible false positive,
 which consists of similar-sized M dwarfs as in Table \ref{tab:res1}. 
 If the false-positive scenarios were true, then we would see the multiple features in the cross-correlation (line profile) of its high-resolution spectrum, 
 since the masses of the two components are similar to each other in the ET scenario.
 In \citet{Hirano2018a}, a single-lined profile was shown for EPIC 220621087, ruling out the false positive scenario that our analysis software has returned.
 Similarly, a PB scenario can be ruled out by AO imaging and cross-correlation profile.
 In addition, phase-curve variations are not seen for both single and double periods in Figure \ref{reflc}, suggesting a low probability with the system being a binary.

The inconsistency of our analysis results made us consider that it is difficult to distinguish a true scenario from other possibilities, when the transit depths for all passbands are similar and relatively shallow ($\lesssim 0.1\%$) in comparison with the precision of the ground-based photometry.
In such cases, 
the scenario with a single star and ones including multiple stars of similar masses 
produce similar observational signals, leading to a mis-classification in the output. 
The same is true in the case of the scenario with small transiting objects
and the scenario hosting grazing object.
Another possible reason for the inconsistency is the incompleteness of the isochrone models; the low-mass range ($< \,0.3\,M_{\odot}$) of the isochrones is known to show systematic deviations in theoretical values from the empirical ones \citep{Mann2015},
suggesting that the analysis for cool targets may have large systematic uncertainties
in the derived parameters.

\subsection{EPIC 220696233}
The most likely scenarios are PS and PB for the MIST isochrones.
In the former case, the central star has a mass of $0.61\,M_{\odot}$, while
for the latter case, the central star is a $0.54\,M_{\odot}$ M-dwarf  
and the companion star has $0.38\,M_{\odot}$.
In both scenarios, the radius of the transiting planet is $\approx 0.06 \,R_{\odot}$.
The flat bottom of the transit curve in Figure \ref{lightcurves} is also suggestive of
the planetary nature of the candidate.
Both analyses with DSEP and PARSEC resulted in PS as the minimum BIC scenario.
The properties in the PS scenarios are consistent within $\approx 2 \sigma$ among these analyses.

As we noted in Section \ref{obs:96233}, no contaminant is resolved 
within $0''.5$ by AO imaging, corresponding to $\approx 130$ au.
In Figure \ref{reflc}, phase-curve variations are not seen at both single and double periods.
In the PS scenario, the planet candidate is a super-Neptune-sized planet orbiting a $0.6\,M_{\odot}$ star (i.e., early M dwarf).
In order to rule out the possibility of EB completely,
high resolution spectroscopy would be required.
Although this target is optically faint ($\sim 16$ mag in the {\it V} band), 
the detection of a $0.5\,M_{\odot}$ - $0.5 \,M_{\odot}$ binary (i.e., the secondary feature in the cross-correlation profile) would be relatively easy
by e.g., near-infrared high-resolution spectroscopy.

 \section{Discussion}\label{sec:discussion}
 \subsection{Isochrone Dependence and Systematic Errors by Adopted Parameters}\label{sec:isochrones}
 There are two types of systematic effects due to the isochrones in the fitting result as shown in Section \ref{sec:results}.
 One pertains to the selection of the most plausible scenario with the minimum BIC value.
 This systematic effect solely depends on the goodness of fit, because the degree of freedom is the same (fixed) for the different isochrones.
In the fitting processes, the light curve model (the first term in Equation \ref{eq:5}) is relatively insensitive to the selected isochrone model, since
only the dilution factor $d_{\lambda}$ depends on the color returned by the isochrone model, and the other orbital parameters are independent of isochrones.
However, the SED model (the second term in Equation \ref{eq:5}) heavily relies on the isochrone-dependent color difference, and
therefore the plausible-model selection by BIC is very sensitive to the isochrones
adopted in the analysis.
The other impact of isochrone models on the fit result is the difference in the derived values for each scenario.
In most of the cases presented in this work, 
the derived values are consistent within $\approx 2 \sigma$ among the three isochrones.
 The bimodal posterior distribution shown by EPIC 206036749 is sometimes seen for ET scenarios, 
 suggesting that there is a degeneracy in the expected combined contrast in cases of more than two stars.
 
 We estimated the expected error propagation from the metallicity uncertainty, adopting EPIC 206036749 as a representative sample. 
 In doing so, we used the MIST isochrones with [Fe/H] = 0.27 and 0.43 and assessed the variations in the derived parameters.
 The best-fit stellar masses in ET are 0.57 - 0.43 -1.06 $\, M_{\odot}$ and 0.60 - 0.46 - 1.06 $M_{\odot}$ for $M_{\rm tra}$ - $M_{\rm cen}$ - $M_{\rm com}$, respectively.
 Thus, the systematic error for the estimated masses arising from fixing the metallicity of the system is no more than $0.03\,M_{\odot}$, which is generally smaller than statistical errors in the derived masses (Table \ref{tab:res1}).  
 We also investigated age-dependency by applying MIST isochrones with 5.0 Gyr,
 finding 0.58 - 0.44 - 1.02 $M_{\odot}$ for $M_{\rm tra}$ - $M_{\rm cen}$ - $M_{\rm com}$.
 The small deviations of those values from the fiducial result (Table \ref{tab:res1})
 suggest that our analysis is insusceptible to relatively small uncertainties in 
 the system age and metallicity, as long as the system is in the main sequence of the
 evolutionary track. 
 
 \begin{table*}[]
\caption{\small The best-fitted parameters for the ``ET" scenario of EPIC 206036749 for each instrument.}\label{tab:res_camera}
\centering
  \begin{tabular}{cccccc} \hline \hline
 Instrument     & $M_{\rm tra}(M_{\odot})$	& $M_{\rm cen}(M_{\odot})$ &$M_{\rm com}(M_{\odot})$	& $\beta$ & $\chi^2_{\rm red}$ \\ \hline
   MuSCAT	& $0.54^{+0.11}_{-0.16}$ & $0.50^{+0.07}_{-0.05}$ & $1.06^{+0.02}_{-0.04}$ & $0.74^{+0.06}_{-0.07}$ & 1.35 \\ 
   SIRIUS		& $0.58^{+0.39}_{-0.08}$  & $0.44^{+0.05}_{-0.04}$ & $1.06^{+0.02}_{-0.37}$ & $0.68^{+0.07}_{-0.08}$ & 1.05 \\ 
  \hline
  \end{tabular}
\end{table*} 

 \subsection{Passband Dependence}
 In following up candidate transiting systems, it is not always possible to conduct multicolor transit photometry from the optical (e.g., by MuSCAT) to near-infrared (e.g., by IRSF) wavelengths as we did for EPIC 206036749 and EPIC 220621087.
 In particular, for the ET scenario containing more than two stars, completely
 different fitting results may be derived, depending on the passbands used in
 the analysis.
 
 To evaluate the dependence of the fitting result on the light-curve passbands used in the analysis, we performed an additional analysis for the ET scenario of EPIC 206036749, which
 shows significant transit-depth variations against wavelength.
 We estimated the properties for the two different subsets of light curves: K2 plus SIRIUS light curves, and K2 plus MuSCAT light curves.
The result of this additional test is shown in Table \ref{tab:res_camera}.
The analysis of K2 plus SIRIUS light curves resulted in bimodal posterior distributions as in Section \ref{ss:36749}, but 
the highest-peak position was consistent with the K2+MuSCAT result within $1 \sigma$.
With regard to the median values, the result for the K2 plus SIRIUS analysis is closer to the result in Table \ref{tab:res1}, in which all the light curves were used.
We conclude that, 
while using a subset of multicolor light curves produces consistent results in most cases, limited light curve data sometimes result in a worse convergence
in the posterior distributions of fitting parameters. 

\section{Conclusion}\label{sec:conclusion}
 In this paper, we have presented a new method to assess the properties of the candidate planetary systems using multicolor photometry:
 the transit/eclipse light curves and apparent magnitudes.
 Based on the observations and analyses for the six K2 systems hosting planetary candidates, we demonstrated the validity and limitation of our method.
 Since this method heavily relies on the isochrone models,
 we have performed the analyses using three different isochrones 
 and confirmed that the derived parameters are almost consistent within 
 $\approx 2 \sigma$ for the three models.
 We also confirmed low passband-related systematic errors in our method for optical (MuSCAT) and near-infrared (SIRIUS) observations.
 We found a disagreement in the low-mass regime of different isochrone models,
 which was already known for the empirical characterizations of low-mass stars \citep[e.g.,][]{Mann2015}.
 This fact suggests that systematic errors in the derived parameters are
 generally large when the plausible scenario involves cool stars less massive
 than $\approx 0.3\, M_{\odot}$.

 While in many cases our method is not able to decisively derive the 
 true scenario, it is still useful to quantitatively constrain the system properties for each scenario, which would be helpful to prioritize the targets in conducting further follow-up observations (e.g., intensive RV measurements).
 Out of the six targets presented in this paper, our analysis implied that all the targets
 exhibiting significant transit-depth variations against wavelength are likely false positives, whereas 
 EPIC 220696233 is most likely a true planetary system, consisting of a super-Neptune-sized planet orbiting an early M dwarf. Such a large planet is relatively scarce around low-mass stars, and therefore further observations would be encouraged.

 Our method will also help the detection of the circumstellar planets in binary systems which cannot be resolved by AO imaging.
 \citet{Berger2018} pointed out a radius inflation for approximately 3,100 cool main-sequence stars in 177,911 Kepler stars of the Kepler Stellar Properties Catalog \citep[KSPC DR25;][]{Mathur2017}, a significant fraction of which could possibly be due to stellar binarity.
 If we assume that those binary systems have planets with a similar occurrence rate as for single stars \citep{Cassan2012},
 they would become key resources to study the structure and dynamical evolution of planets in multiple-star systems.
Given the observational efficiency, 
our technique to use multicolor photometry would be very useful
in validating the planet candidates in those binary systems, which is the first step for such studies. 

\acknowledgments

 This paper is based on data collected at IRSF and Okayama 188cm telescope.
 We thank Katsuhiro Murata, Takahiro Nagayama and the other members of IRSF team 
 for support of the observation.
 This work was supported by Japan Society for Promotion of Science (JSPS) KAKENHI 
 Grant Numbers JP19J21733, JP17H04574,  and JP19K14783, JP18H01265 and JP18H05439,
 and JST PRESTO Grant Number JPMJPR1775. 
 This paper uses observations made at the South African Astronomical Observatory (SAAO).
 This work has made use of data from the European Space Agency (ESA) mission
{\it Gaia} (\url{https://www.cosmos.esa.int/gaia}), processed by the {\it Gaia}
Data Processing and Analysis Consortium (DPAC,
\url{https://www.cosmos.esa.int/web/gaia/dpac/consortium}). 
Funding for the DPAC has been provided by national institutions, in particular the institutions participating in the {\it Gaia} Multilateral Agreement.
 \bibliographystyle{aasjournal}
 \bibliography{miyakawa2019.bib}
\end{document}